\documentclass[twocolumn,showpacs,preprintnumbers,amsmath,amssymb]{revtex4}

\usepackage{dcolumn}
\usepackage{bm}
\usepackage{epsfig}
\usepackage{graphicx,psfrag}

\newcommand{\be}{\begin{equation}}
\newcommand{\ee}{\end{equation}}
\newcommand{\bear}{\begin{eqnarray}}
\newcommand{\eear}{\end{eqnarray}} \newcommand{\ba}{\begin{array}}
\newcommand{\ea}{\end{array}}

\newcommand{\gae}{\begin{array}{c}\,\sim\vspace{-1.7em}\\> 
\end{array}}

\def\beq{\begin{equation}}
\def\eeq#1{\label{#1}\end{equation}}
\def\eeqn{\end{equation}}
\def\eeq{\end{equation}}
\def\beqa{\begin{eqnarray}}
\def\eeqa#1{\label{#1}\end{eqnarray}}
\def\eeqan{\end{eqnarray}}

\def\to{\rightarrow}

\newcommand\iden{\leavevmode\hbox{\small1\normalsize\kern-.33em1}}

\def\W3{W_H^3}

\begin{document}

\title{Direct Detection of Supersymmetric Particles in Neutrino Telescopes}
\author{Ivone~F.~M.~Albuquerque and 
Gustavo Burdman }
\affiliation{Instituto de F\'isica, Universidade de S\~ao Paulo, 
S\~ao Paulo, Brazil}
\author{Z. Chacko}
\affiliation{Department of Physics, University
of Arizona, Tucson, AZ~85721, USA}
\vspace*{0.3cm}


\begin{abstract}

In supersymmetric theories where the lightest supersymmetric particle is the
gravitino the next to lightest supersymmetric particle is typically a long lived
charged slepton.  In this paper, following our earlier proposal~\cite{abc}, 
we perform a detailed study of the production of
pairs of these particles induced by the interactions of high energy cosmic neutrinos
with nucleons in the earth, their propagation through the earth and finally their
detection in neutrino telescopes. We investigate the charged slepton energy loss in
detail and establish that the relatively small cross-section for the production of
supersymmetric particles is partially compensated for by the very long range of these
heavy particles.  The signal, consisting of two parallel charged tracks emerging from
the earth, is characterized by a track separation of a few hundred
meters. We perform a careful analysis of the main background, coming from
direct di-muon production, and show that it can be separated from the signal due to
its characteristically smaller track separation. We conclude that neutrino telescopes
will complement collider searches in the determination of the supersymmetry breaking
scale, and may even provide the first evidence for supersymmetry at the weak scale.

\end{abstract}

\maketitle

\section{Introduction}
\label{intro} 

One of the most pressing questions in particle physics is the origin and stability of
the hierarchy between the weak and the Planck energy scales. Natural solutions of
this so called hierarchy problem require new physics at the TeV scale.  One of the
most attractive candidate theories is weak scale supersymmetry. Although this is in
no small measure due to its theoretical appeal (it is a simple and natural extension
of the usual space-time symmetries), it is also favored by data from electroweak
observables. These point to a weakly coupled Higgs sector, one without significant
deviations from the standard model in regard to electroweak precision observables, as
arises naturally in supersymmetric theories. However, supersymmetry must be broken
since the superpartners have not yet been observed. The supersymmetric spectrum is
determined by the supersymmetry breaking mechanism.\\ 
\indent 
Supersymmetric
theories typically possess a discrete symmetry, R-parity, which ensures that 
corrections to precision electroweak observables are small. The existence of this
discrete 
symmetry immediately implies 
that the 
Lightest Supersymmetric Particle (LSP) is
stable. 
Which of the supersymmetric particles is the LSP is determined by the scale
of supersymmetry breaking, which we denote by $\sqrt{F}$.
When supersymmetry is broken at high scales
such that $\sqrt{F} \gtrsim 10^{10}$GeV the LSP is typically the 
neutralino.
If however supersymmetry is broken at lower scales, $\sqrt{F} \lesssim 10^{10}$GeV, 
the LSP
tends to be the gravitino. In models where the LSP is the gravitino, the Next to
Lightest Supersymmetric Particle (NLSP) is usually a charged slepton, typically the
right-handed stau. When the supersymmetry breaking scale $\sqrt{F}$ is much
larger than a TeV
the stau NLSP 
decays to gravitinos through interactions that are
extremely small, and therefore
its lifetime can be very large. 
In gauge-mediated SUSY breaking, for instance, we have
\beq
 c\tau = \left(\frac{\sqrt{F}}{10^7{\rm~GeV}}\right)^4\, 
\left(\frac{100~{\rm GeV}}{m_{\tilde\tau_R}}
\right)^5\,10~{\rm km}~,
\label{ctau}
\eeqn
where ${m_{\tilde\tau_R}}$ is the stau mass.  Thus, for $\sqrt{F} \gtrsim 10^7$GeV if
these NLSPs were to be produced by very high energy collisions they could travel very
long distances before decaying. 

Many interesting and realistic supersymmetric models have been proposed where the
scale of supersymmetry breaking $\sqrt{F}$ is larger than $5 \times 10^6$GeV and
where the LSP is the gravitino while the NLSP is a long-lived stau. These include
models of low energy supersymmetry breaking such as gauge mediation{\cite{GMSB}}, but
also superWIMP scenarios{\cite{superWIMP}}, where the supersymmetry breaking scale is
much higher and the NLSPs decay to the gravitino only on very large time scales of
about a year. 

In a recent letter~\cite{abc} we proposed that the diffuse flux of high
energy neutrinos colliding with the earth could produce pairs of slepton NLSPs which
due to their large range could travel large distances through the earth and be
detected in neutrino telescopes.  The neutrino-nucleon reaction can produce a pair of
supersymmetric particles that will promptly decay to a pair of NLSPs and standard
model (SM) particles. This process results in NLSPs which typically have a very high boost
and therefore will not decay inside the
earth provided the supersymmetry breaking scale $\sqrt{F} > 10^7$GeV. (For $5 \times
10^6 < \sqrt{F} < 10^7$GeV a significant fraction of the decays will occur inside the
earth.) Since the NLSP is charged, its upward going tracks could in principle be
detected in large ice or water Cerenkov detectors, such as IceCube~\cite{icecube}.  
This is in analogy with the standard model charged current interaction giving muons,
the primary signal in neutrino telescopes. The high boost and large range of the
NLSPs implies that the tracks are parallel and well separated, enabling them to be
distinguished from background events. Various aspects of this scenario were further
considered in Refs.{\cite{bi,reno3,ring}}.

In principle one could think that cosmic rays could be an even better source
of supersymmetric events, and ultimately of NLSPs. After all, if one considers a proton 
primary, squark pair production should be larger than the slepton production via
neutrino interactions, due to the fact that it goes through a larger coupling. 
However, this cannot compete with the soft, forward cross section of cosmic rays 
with air, which is large 
enough to stop primaries in the atmosphere.  
Typical cross sections for cosmic ray primaries in the earth's atmosphere are around
$10-100$~mb, so that the atmosphere is several interactions lengths.  
On the other hand, supersymmetric production cross sections at the energies of 
interest are in the order of $10-100~$pb, or $10^{-9}$ times smaller. 
Thus, the probability of
producing supersymmetric events with proton or other cosmic ray primaries is very small. 
Neutrinos, on the other hand, 
will always go through the atmosphere, with enough of them
interacting inside the earth within the range of the NLSP from the detector.
Then, since the neutrino flux is not expected to be that much smaller than the 
cosmic ray flux, using neutrinos is advantageous.

In this paper we present a detailed analysis of the original proposal~\cite{abc}. 
We perform a
careful study of the production of supersymmetric particles, their propagation
through the earth, and their detection in neutrino telescopes. We compute the
expected number of signal events in IceCube, their characteristic energy
distribution and the energy deposited in the detector. We also investigate the main
background, which arises from di-muon production, and verify that it can separated
from the signal by making suitable cuts.

In analyzing the signal, it is particularly important to carefully incorporate the
energy losses involved in the propagation of sleptons through the earth. Although
Bremsstrahlung and pair-production energy loss become less important for heavy
charged particles, photo-nuclear energy loss potentially presents a problem. In fact,
at first this appears to be a great obstacle since it is known that photo-nuclear
energy loss dominates the propagation of tau leptons~\cite{tau1,tau2}.  However, this
is not the case. As we already sketch in Ref.~\cite{abc} and was shown by the
explicit calculation in Ref.~\cite{reno3}, there is a mass suppression of the
photo-nuclear energy loss, which manifests itself for masses well above that of the
$\tau$ and affects the energy loss of a particle such as the $\tilde\ell_R$.  We will
reproduce the results of Ref.~\cite{reno3} regarding the photo-nuclear energy loss of
a slepton, and explicitly show how the mass suppression comes about and why it is not
operative for the $\tau$ energy loss.  As a result the photo-nuclear energy losses 
for
heavy charged particles, whether scalars such as the NLSP or fermions such as heavy
leptons, are suppressed due to their large masses.

The main background to the signal arises from di-muon events which also give rise to
pairs of tracks in the detector.  However, we show that these can be readily
distinguished from the signal by the separation between tracks. The key point is that
the NLSP range is much larger than the muon range, which is only of order a few
kilometers.  Therefore while the typical track separation in the detector for the
NLSP signal is of order a few hundred meters, direct di-muon production results in
tracks that are typically closer together by an order of magnitude.

The plan for the paper is as follows: we review in detail the relevant cross section
for supersymmetric production in Section~\ref{xsecs}.  In Section~\ref{dedx} we study
in detail the energy loss of sleptons as they travel through the earth. An analysis
of the signal and backgrounds is presented in Section~\ref{signal} and we conclude
with a discussion of our results in Section~\ref{conc}.

\section{Cross Sections}
\label{xsecs}
\begin{figure}
\centering
\epsfig{file=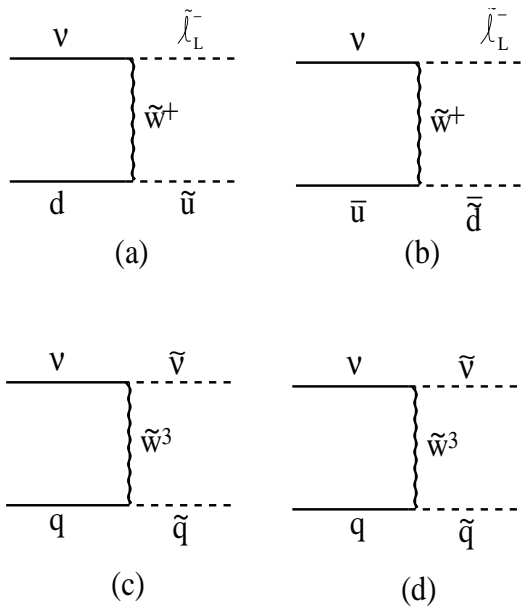,width=7cm,height=8cm,angle=0}
\caption{
Feynman diagrams for supersymmetric particle production in $\nu N$ collisions 
Charged current (chargino) interactions: (a) Left-left interaction requiring
the insertion of the gaugino mass in the t-channel line. 
(b) Left-right interaction.  
Neutral current: (c), (d). 
There are analogous diagrams for anti-neutrinos as well as for 
strange and charm initial quarks. 
}
\label{feynman}
\end{figure}

In this section we summarize the calculation of the production cross section 
for the NLSP pair. The supersymmetric process of interest 
involves the t-channel production of a slepton and a squark via 
gaugino exchange. For simplicity, we neglect mixing with Higgsinos 
in the gaugino sector. Then the processes analogous to the charged 
current (CC) in the SM is $\nu N\to \tilde{\ell}_L \tilde{q}$, where 
$\tilde{q}$ can be an up or down-type squark, and the particle
exchanged in the t-channel is a chargino. These are shown in 
Figure~\ref{feynman}(a) and (b), and are the dominant contributions.  
The neutrino, always produced
left-handed by the 
weak interactions, can interact either with a left-handed down-type 
quark (a), or with a 
right-handed up-type anti-quark (b). 
This results in the partonic cross sections:
\beqa
\frac{d\sigma^{\rm (a)}}{dt} &=& \frac{\pi\alpha}{2\,\sin^4\theta_W}\,
\frac{M^2_{\tilde w}}{s\,(t-M^2_{\tilde w})^2}
\label{llcc}\\
\frac{d\sigma^{\rm (b)}}{dt} &=& \frac{\pi\alpha}{2\,\sin^4\theta_W}\,
\frac{(tu-m_{\tilde\ell_L}^2\,m_{\tilde q}^2)}{s^2\,(t-M^2_{\tilde w})^2}
\label{lrcc},
\eeqan
where $s,t$ and $u$ are the usual Mandelstam variables, and 
$M_{\tilde w}$, $m_{\tilde\ell_L}$ and $m_{\tilde q}$ are the chargino, 
the left-handed slepton and the squark masses respectively. 
The left-handed slepton and the squark decay promptly to the 
lighter ``right-handed'' slepton 
plus 
non-supersymmetric particles. 
We also include the subdominant neutralino exchange, 
Figure~\ref{feynman}~(c)-(d). 
\begin{figure}[t]
\centering\leavevmode \epsfxsize=325pt \epsfbox{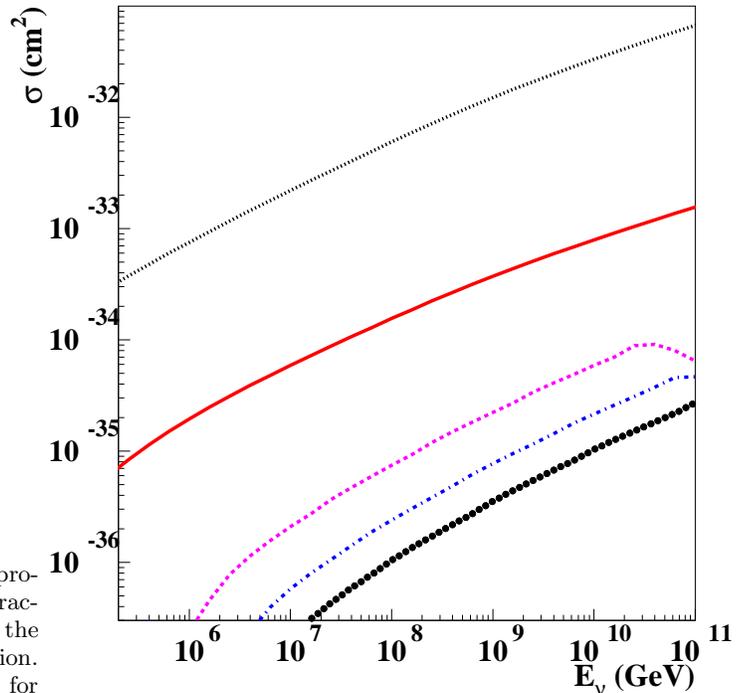}
\caption{$\nu N$ cross sections vs. the energy of the incident neutrino.
The three lower curves correspond to $m_{\tilde\ell_L}=250~$GeV, $m_{\tilde w}=250$GeV;
and for squark masses  
$m_{\tilde q}=300$~GeV (dashed) ,~$600$~GeV (dot-dashed) and $900~$GeV (dotted).
The top curve corresponds to the SM charged current interactions and the middle one
to the di-muon background. 
}
\label{fig1}
\end{figure}   

We take $m_{\tilde w}=250$~GeV, $m_{\tilde\ell_L}=250$~GeV and three 
values for the   
squark masses: $m_{\tilde q}= 300,~600$ and $900$~GeV. These are very 
representative 
values in the scenarios under consideration. 
This is clearly the case for gauge mediated supersymmetry breaking,
where super-partner masses are generated by their interactions.  
Then, typically the
$\tilde\tau_R$ 
is the NLSP, being
heavier only than the ultra-light and very weakly coupled gravitino. 
Charginos and neutralinos tend to be heavier since
they also feel the $SU(2)_L$ interactions. Finally, squarks are
heavier 
still since their masses are
affected by the strong interactions. 
In Figure~\ref{fig1} we plot the cross sections for supersymmetric 
production in $\nu N$ interactions
as a function of the neutrino energy. Also plotted for comparison is
the SM charged current (top curve) the di-muon (second from top) 
cross sections.

Every supersymmetric event produces a 
$\tilde\ell_L$ and $\tilde q$ as primaries, 
which then decay. The resulting cascades always leave a pair of $\tilde\ell_R$ 
and SM particles.
In what follows we will take $m_{\tilde\ell_R}=150$~GeV.
The SUSY cross sections are considerably suppressed
with respect to  the SM, even when well above threshold. 
This is to be expected since the SM cross section is dominated by very
small values of $x$, whereas for the SUSY processes one always needs
to be above a rather large threshold, resulting in $x > m^2_{\rm
SUSY}/(2 M_P E_\nu)$, where $m_{\rm SUSY}$ is the typical mass of the 
supersymmetric particles being produced. 
However, we will see
that this will be compensated by the very long range of sleptons
compared with the muon range, which is only a few kilometers. 

\section{NLSP Energy Loss}
\label{dedx}

Once produced by the $\nu N$ interactions in the earth, the  pair of 
NLSPs should 
range into the detector, just as the muons produced by CC events~\cite{als}. 
Charged particles lose energy due to ionization processes as well as 
through radiation.
The average energy loss can be written as~\cite{pdg}
\beq
-\frac{dE}{dx} = a(E) + b(E)~E~,
\label{eloss}
\eeq
where $a(E)$ characterizes the ionization losses, and $b(E)$ includes 
the contributions to 
radiation energy losses from various sources including bremsstrahlung,
pair production 
and photo-nuclear interactions. 

\subsection{Ionization}
Ionization energy loss can be approximated by the 
Bethe-Block formula~\cite{pdg}
\bear
\left(-\frac{dE}{dx}\right)_i & = & 2\pi N_A\,r_e^2m_ec^2\,
\frac{z^2}{\beta^2}\,\frac{Z_a}{A_a}\;
\times\nonumber\\
& & \left\{\ln\left(\frac{4m_e^2c^2v^2\gamma^2(\beta\gamma)^2}{I^2}\right)
-2\beta^2 \right.\nonumber\\
& & \left.- \delta -\frac{C}{Z_a}\right\}~.
\label{ion}
\eear
In eqn.(\ref{ion}), $N_A$ is Avogadro's number, $r_e$ is the classical
electron radius, $m_e$ its
mass, $Z_a$ and $A_a$ the atomic number and mass of the absorber
respectively, and $z$ the charge of 
the incident particle in units of $e$, the electron charge. 
The mean ionization potential $I$ is 
typically of about $100~$eV ($136~$eV for Standard Rock). The density 
correction $\delta$ reflects
the effect of the polarization of the medium as the incident
particle's energy increases. This reduces
the energy loss at high energies. For the case of the NLSP in
question, this reduction is still 
small and of the order of $10\%$. Finally, the shell correction $C$
taking into account the effects
of atomic bounding, are only important for small values of
$\beta\gamma$, 
similar to the 
bound electrons'. For $\beta\gamma\simeq0.3$ this correction is about $1\%$. 
For the much larger values of $\beta\gamma$ relevant here, this
correction is 
negligible. 
Therefore, the  ionization loss can be parameterized by 
\be
a(\beta\gamma) \simeq   0.08~\frac{\rm MeV~cm^2}{\rm gr}\,\left(17 + 
2\ln\beta\gamma\right),
\label{ae}
\ee
and is rather independent of the particle mass.  For instance, for a 
density of 
$\rho = 2.5~{\rm gr}/{\rm cm}^3$, roughly corresponding to standard
rock in the earth's crust, 
and for $\beta\gamma=10^5$, we have $a\simeq 8~{\rm MeV}/{\rm cm}$.
However, at these values of $\beta\gamma$ most of the energy loss will
occur through radiation. 

\subsection{Radiation Energy Loss}
At sufficiently large energies the energy loss due to radiation
processes dominates
over ionization for all charged particles. For instance, for muons
this dominance takes place at energies of several hundred GeV. 
The radiation loss can be written as 
\beq
b(E) = \frac{N_A}{A}\int_{y_{min}}^{y_{max}} y\, \frac{d\sigma}{dy}\, dy~,
\label{bgeneric}
\eeq
where $N_A$ is Avogadro's number, $A$ is the atomic mass of the
target, and $y$ is the fractional energy loss of the (s)lepton
\beq
y \equiv \frac{E-E'}{E}~,
\label{ydef}
\eeq
with $E'$ the final lepton energy in the target's rest frame.
Although there is no simple scaling with mass, heavier particles 
such as NLSPs will have suppressed radiative energy losses. 
In what follows we discuss the differential cross section $d\sigma/dy$
for radiative energy loss of NLSPs due to 
Bremsstrahlung, Pair Production and Photo-nuclear interactions.

\subsubsection{Bremsstrahlung}
The differential cross section for Bremsstrahlung is reviewed in detail in
Reference.~\cite{groom}. The contribution from the interaction with the screened nucleus
dominates and is given by
\beq
\left(\frac{d\sigma}{dy}\right)_{brem.,nucl.} = \alpha
\left(2Z\frac{m_e}{m_\ell} 
r_e\right)^2\,
\left(\frac{4}{3} - \frac{4}{3}y+y^2\right)\,\frac{\Phi(\delta)}{y}~,
\label{brem_nucl}
\eeq
where $m_\ell$ is the mass of the interacting (s)lepton, 
\bear
\Phi(\delta) &=& \ln\left(\frac{B m_\ell
Z^{-1/3}/m_e}{1+\delta\sqrt{e} 
B Z^{-1/3}/m_e}\right)\nonumber\\
&&-\ln\left(\frac{D_n}{1+\delta(D_n\sqrt{e}-2)/m_\ell}\right)~,
\label{phidef}
\eear
and we have defined 
\beq
\delta = \frac{m_\ell y}{2 E(1-y)}~,\nonumber
\eeq
and 
\be
D_n = 1.54 A^{0.27}~.
\ee
Finally, $r_e$ is the classical electron radius and in this case $B=182.7$.
There is also a subdominant contribution from the Bremsstrahlung
induced by atomic 
electrons. 
This is down by a factor of $Z$ with respect to
eqn.(\ref{brem_nucl}). It is 
possible to approximate
this effect by replacing $Z^2$ in eqn.(\ref{brem_nucl}) by $Z(Z+1)$. 
In order to obtain $b_{brem.}(E)$ we must integrate $y$, with 
$y_{min}=0$ and $y_{max}= 1-(3/4)\sqrt{e}\,Z^{1/3}\,(m_\ell /E)$.

\begin{figure}
\centering
\epsfig{file=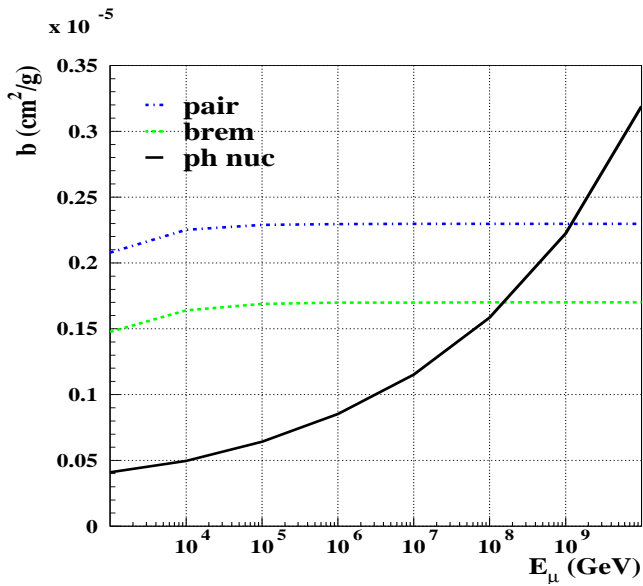,width=8cm,height=8cm,angle=0}
\caption{The muon energy loss function $b(E)$  defined by eqns.~(\ref{eloss})
and (\ref{bgeneric}).  
}
\label{bmu_fig}
\end{figure}

As it can be seen by comparing Fig.~(\ref{bmu_fig}) and 
Fig.~(\ref{bstau_fig}), 
the Bremsstrahlung contributions 
are greatly suppressed for the slepton case compared to the 
muon. 
This can already be seen in the expression
(\ref{brem_nucl}), 
by the appearance of a
factor $1/m_\ell^2$. Thus, for the NLSP case, with masses in the
hundreds of GeV, this contribution will be negligible. 

\subsubsection{Pair Production}
As was noted in Ref.\cite{reno3}, the dominant contributions
to energy loss due to pair 
production come from diagrams which are independent of the spin of the
heavy lepton. Then it is possible
to accurately estimate this contribution to $b(E)$ by considering a 
heavy lepton, instead of a scalar lepton
as it would be the case for the NLSP. The fermion and scalar cases  
are almost identical at the high energies considered here.
A compete QED calculation of the direct pair production was done in 
Ref.\cite{kelner}. Based on this, a useful parametrization for the
double differential cross section is given in Ref.\cite{kokoulin}:
\be
\frac{d^2\sigma}{dy d\rho} =  \alpha^4\, (Z\lambda_e)^2
\,\frac{1-y}{y}\,\left(\phi_e + \frac{m_e^2}{m_\ell^2}\phi_\mu~\right),
\label{dsdypair}
\ee
where $\lambda_e$ is the electron's Compton wavelength and
$\rho \equiv (E^+-E^-)/(E^++E^-)$ is the asymmetry parameter of the 
pair. The functions $\phi_e$ and $\phi_\mu$ correspond to different
QED diagrams and include corrections for atomic and nuclear
form-factors. Their form can be found, for instance, in Reference~\cite{groom}. 
Just as in the Bremsstrahlung case, and in order to take into account the 
contribution from atomic electrons, we can simply replace $Z^2$ in
eqn.(\ref{dsdypair}) by $Z(Z+1)$. Then, the energy loss due to pair
production is computed as 
\be
b_{pair}(E) =
\frac{N_A}{A}\int_{y_{min}}^{y_{max}}\int_{0}^{\rho_{max}}
y\,\frac{d^2\sigma}{dy d\rho}\, dy d\rho ~.
\label{bpair}
\ee 
Here, the integration limits are
\bear
 \frac{4m_e}{E} \leq &y& \leq
1-\frac{3}{4}\sqrt{e}\,Z^{1/3}\,\frac{m_\ell}{E}
\nonumber \\
&{\rm and} \label{limits_pair}\\
0\leq &\rho & \leq
\left(1-\frac{6m_\ell^2}{(1-y)E^2}\right)\,\sqrt{1-\frac{4m_e}{y\,E}}
\nonumber
\eear
The energy loss by $e^+-e^-$ pair production for a muon
is plotted in Fig.~(\ref{bmu_fig}) and that 
for a slepton with $m_{\tilde{\ell}}=150$~GeV in Fig.~(\ref{bstau_fig}) 
(dot-dashed curves).
Once again, it is clear that the energy loss is suppressed
in the case of the heavier (s)lepton. Although pair creation is an
important source of energy loss in the slepton propagation, it is not
the most important one, as we will see next.

\subsubsection{Photo-nuclear Energy Loss}
For a heavy particle such as the slepton NLSP, the
most important source
of energy loss is the photo-nuclear interactions. 
These already dominate the
$\tau$ energy loss. However, as we will see below, there will still be
a considerable suppression 
due to the NLSP mass. This mass dependence is subtle, so we 
spend some time discussing its origin.
\begin{figure}
\centering
\epsfig{file=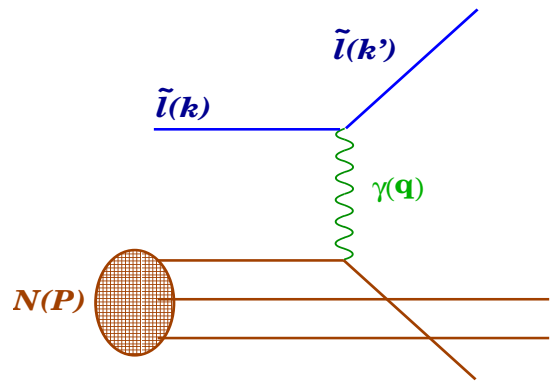,width=8cm,height=5cm,angle=0}
\caption{The scattering $\tilde{\ell}(k) N(P)\to \tilde{\ell}(k') X$ 
in the deep inelastic picture.
}
\label{dis_fig}
\end{figure}   
We compute the scattering of a scalar such a the $\tilde{\ell}$ NLSP 
off a nucleon following the 
deep inelastic formalism. In general the cross section can be expressed as
\be
d\sigma = \alpha^2 \,\frac{1}{EE' Q^4}\,L_{\mu\nu}\,W^{\mu\nu} d^3k'~,
\label{dsigmapn}
\ee
where $E (E')$ is the incoming (outgoing) electron energy, $\vec{k'}$
is the 
outgoing slepton momentum, 
and $Q^2=-q^2=(k-k')^2$ is the photon virtuality, $k$ and $k'$ are
the four-momenta of the incoming and outgoing slepton respectively. 
The momentum fraction of the struck quark is described by the Bjorken 
variable
\be
x = \frac{Q^2}{2P.q} = \frac{Q^2}{2M(E-E')}~,
\ee
with $P$ and $M$ the four-momentum and mass of the nucleon
respectively. 
The hadronic tensor is
parametrized as usual in terms of 
two structure functions, $F_1(x,Q^2)$ and $F_2(x,Q^2)$:
\bear
W_{\mu\nu} &=& -\frac{F_1}{M}\,g_{\mu\nu} + \frac{F_2}{M^2Ey}P_\mu P_\nu
+\left(\frac{F_1}{M} + F_2\frac{Ey}{q^2}\right)\frac{q_\mu q_\nu}{q^2}
\nonumber\\
&& -\frac{F_2}{M}\frac{P_\mu q_\nu + P_\nu q_\mu}{q^2}~.
\label{wmunu}
\eear
In the Callan-Gross limit, the hadronic form-factors are related by 
\be
2xF_1(x,Q^2) = \left(1 + \frac{4M^2x^2}{Q^2}\right) F_2(x,Q^2)~.
\label{callangross}
\ee
More generally, departure from this limit, can be parametrized by the function 
$R(x,Q^2)$ defined by 
\be
R(x,Q^2) = \frac{F_L(x,Q^2)}{2x\,F_1(x,Q^2)}~,
\ee
where the longitudinal structure function is
\be
F_L(x,Q^2) = \left(1 + \frac{4M^2x^2}{Q^2}\right) F_2(x,Q^2) - 2xF_1(x,Q^2)
\ee
Thus, we can eliminate $F_1(x,Q^2)$ in favor of $F_2(x,Q^2)$ and
$R(x,Q^2)$, 
where the latter function vanishes in the deep inelastic limit. 

On the other, hand the slepton tensor is given by 
\be
L_{\mu\nu} = (k+k')_\mu (k+k')_\nu~.
\label{lmunu}
\ee
We can now compute the double differential cross section in terms of 
the structure function
$F_2(x,Q^2)$ as well as the Callan-Gross violating function
$R(x,Q^2)$. 
The result is given by 
\bear
\frac{d^2\sigma}{dy dQ^2} &=&  \frac{\pi\alpha^2}{Q^4}\,\frac{F_2(x,Q^2)}{y}
\left\{(2-y)^2 \right.\nonumber\\
&& \left. - y^2\left(1+\frac{4m_\ell^2}{Q^2}\right)\,
\frac{\left(1+\frac{4M^2x^2}{Q^2}\right)}{(1+R(x,Q^2))}
\right\}~.
\label{d2syq2} 
\eear
which agrees with Ref.~\cite{reno3}. 
In order to obtain the photo-nuclear energy loss function we have 
\be
b_{\rm ph-nuc}(E) = \frac{N_A}{A} \int_{y_{\rm min}}^{y_{\rm max}}dy\,
\int_{Q^2_{\rm min}}^{Q^2_{\rm max}} y\,\frac{d\sigma}{dy dQ^2}~,
\label{bpnuc}
\ee
where $y_{\rm min} = ((M+m_\pi)^2-M^2)/2ME$, 
$y_{\rm max}= 1-m_{\tilde\ell}/E$, 
$Q^2_{\rm max} = 2MEy-((M+m_\pi)^2 - M^2)$ and the minimum photon 
virtuality is given by
\be
Q^2_{\rm min} = 2\left[E^2(1-y) - k k' - m_{\tilde\ell}^2\right]\simeq 
\frac{m_{\tilde\ell}^2\,y^2}{(1-y)}
~,\label{q2min}
\ee
and corresponds to a forward slepton. The last expression 
in eqn.(\ref{q2min})
reflects the approximation for small values of the slepton energy 
loss ($y\ll 1$).
\begin{figure}
\centering
\epsfig{file=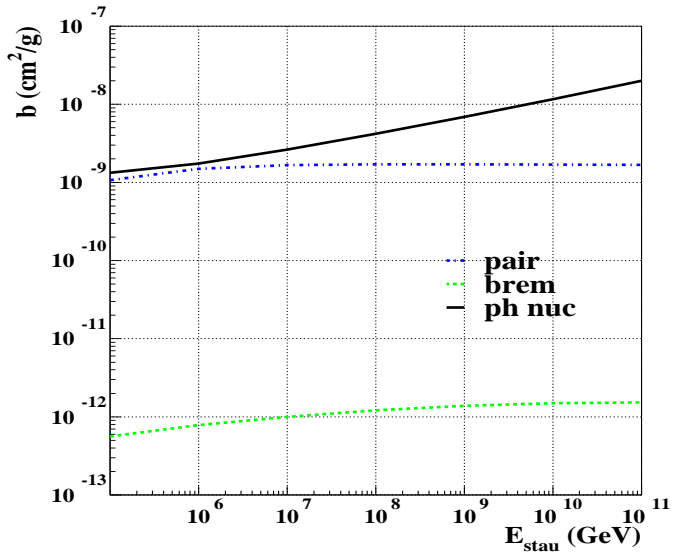,width=8cm,height=8cm,angle=0}
\caption{The slepton energy loss function $b(E)$  defined 
by eqns.~(\ref{eloss})
and (\ref{bgeneric}).  
}
\label{bstau_fig}
\end{figure}   
The Callan-Gross limit is an excellent approximation in the energies
of interest 
here, so we will take $R(x,Q^2)=0$ from now on. We must now specify
the structure function $F_2(x,Q^2)$. We use the ALLM 
parametrization~\cite{allm1,allm2},
which fits nicely all data from very low $Q^2$ to $Q^2=5000~{\rm
GeV}^2$. The parametrization has a $\chi^2/{\rm ndf} =0.97$ for a
total of 1356 points. The details of the parametrization used here 
can be found in the update of Ref.~\cite{allm2}. 

In addition, we must account for nuclear shadowing, which can be an
important  effect at low values of $x$.  This is done by defining
the ratio
\be
a(A,x,Q^2) = \frac{F_2^{A}(x,Q^2)}{AF_2^N(x,Q^2)}~,
\ee
where $F_2^{A}(x,Q^2)/{A}$ and $F_2^N(x,Q^2$ are the nuclear  and
the free nucleon structure functions respectively. A parametrization
of this ratio describing the data is given in Ref.~\cite{tau1}. 
It is  found to be rather $Q^2$-independent and is given by
\bear
a(A,x) = \left\{
\begin{array}{cc}
A^{-0.1} & x<0.0014 \\
A^{0.069\log_{10}+0.097} & 0.0014\leq x \leq 0.04 \\
1                        &  0.04 < x
\end{array}
\right.
\nonumber
\eear
Assuming that $Z=A/2$, the structure function $F_2^{A}$ can be obtained as
\be
F_2^{A} = a(A,x) \frac{A}{2}\left(1+P(x)\right) F_2^p~.
\label{shadowing}
\ee
In eqn.(\ref{shadowing}) $F_2^p$ is the proton structure function and 
$P(x) = 1-1.85x+2.45x^2 - 2.35x^3 +x^4$ describes the $F_2^n/F_2^p$ ratio as
a fit to the BCDMS data. 

The photo-nuclear energy loss can now be computed using
eqn.~(\ref{bpnuc}) with  
the structure function $F_2^p(x,Q^2)$ that is extracted from data  
and parametrized by ALLM. 
We plot the result in Fig.~\ref{bstau_fig} (solid curve). 

We observe that the
photo-nuclear energy loss appears to be also suppressed for higher
masses when compared with the muon case, as can be seen from
Fig.~\ref{bmu_fig} (solid curve).
This seems odd at first, since the effect is larger for the $\tau$
than it is for the $\mu$~\cite{tau1,tau2}. 
In order to understand this mass dependence for the heavier masses, we
should note that the cross section that determines the structure 
function $F_2^p(x,Q^2)$ 
is dominated by physics at rather low $Q^2$ ($\simeq 1~{\rm GeV}^2$),
where resonances and other non-perturbative effects set in.
However, we see from eqn.~(\ref{q2min}) that for large slepton masses
the minimum value of the photon virtuality is typically larger than
this, unless the energy loss $y$ is very small. Thus, the most
important contributions to the structure function are left out as a
consequence, leading to a suppression for sufficiently large masses
$m_{\tilde\ell}^2 \gg 1{\rm GeV}^2$. This explains why the effect is still 
large for the $\tau$ but then is suppressed for heavier (s)leptons.

Finally, we consider the possible electroweak contributions to the slepton energy
loss, where the photon is replaced by either a gauge boson or a gaugino. 
Since the propagating slepton is ``right-handed'', it only  has neutral interactions.
Of these, the dominant contribution comes from the exchange of a $Z$ boson. For this 
to begin to be comparable to the photo-nuclear process described above, 
we need $Q^2\gae M_Z^2$. However, as mentioned earlier, the structure functions 
are dominated by the low $Q^2$ region. Thus, we expect the $Z$-exchange contribution to be 
suppressed by this effect. The results of the explicit calculation confirms this expectation.
The energy loss due to $Z$-exchange is at least two orders of magnitude smaller than the 
photo-nuclear energy loss, and even smaller for most of the energies
of interest. As a source 
of energy loss it is also smaller than the pair-production, 
and is only larger than bremsstrahlung
for slepton energies above $10^{7}$~GeV.

In what follows we will make use of the calculations of the  
energy loss for an NLSP in order to compute the number of events 
ranging into the IceCube detector.

\section{Signals in Neutrino Telescopes} 
\label{signal}

\subsection{Neutrino Flux}

In order to compute the event rates in neutrino telescopes, we need to
know the incoming neutrino flux. The presence of cosmic neutrinos is 
expected on the basis of the existence of high energy cosmic rays. 
Several estimates of the neutrino flux  are available 
in the literature. In most cases, it is expected that km$^3$ 
neutrino telescopes
will measure this flux. Here we will make use of two calculations 
of the neutrino flux in order to compute the number of expected 
NLSP events at neutrino telescopes. 

First, we consider the work of Waxman and Bahcall (WB)~\cite{wb}, who
pointed out that the observed cosmic ray flux implies an upper bound
on the high energy astrophysical neutrino flux. This argument
requires that the sources be optically ``thin'', meaning that most of
the protons escape and only a fraction of them interact inside the
source. In order to determine the neutrino spectrum, WB fix the 
cosmic ray spectral index  to $-2$, giving
\be
\left(\frac{d\phi_\nu}{dE}\right)_{\rm WB} = \frac{(1-4) \times
10^{-8}}{E^2} {\rm GeV~ cm^{-2} s^{-1} ~sr{-1}}~,
\label{wblimit}
\ee
where the range in the coefficient depends on the cosmological
evolution of the sources. Using the upper end we obtain the so
called
WB limit.

On the other hand, Manheim, Proterhoe and Rachen (MPR)~\cite{mpr}
obtain and upper limit on the diffuse neutrino sources in a way that 
is somewhat different than WB. Instead of assuming a fixed cosmic ray
index for the spectrum, MPR determine the spectrum directly from data
at each energy. In the case when the sources are considered optically
``thin'' this procedure results in a limit similar to that of WB
for energies between $10^7$ and $10^9$ GeV and larger otherwise.

We consider an initial flux containing both 
$\nu_\mu$ and $\nu_e$~(in a 
$2:1$ ratio). Since the 
initial interactions (see Figure~\ref{feynman}) produce $\tilde\ell_L$
and these are nearly degenerate 
in flavor, the flavor of the initial neutrino does not affect our results. 
For the same reason, the possibility of large mixing in the neutrino 
flux is also innocuous here.

\subsection{NLSP Signals}
Having the neutrino flux from the previous section, 
the production cross section for NLSP pairs 
(Section~\ref{xsecs}),
and their energy loss (Section~\ref{dedx}) through earth, 
we can now compute the number of 
NLSP events at neutrino telescopes. 
In order to correctly take into account the propagation of neutrinos 
and the NLSP $\tilde\ell_R$ through the earth, we make use of a model 
of the earth density profile  as detailed
in Reference~\cite{gqrs}. 

\begin{figure}
%
\begin{flushright}
\epsfig{file=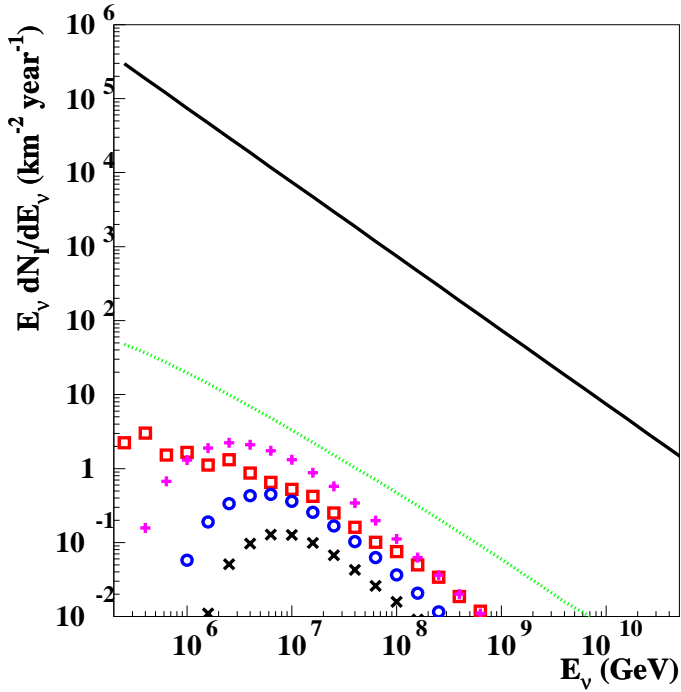,width=11.2cm,height=11cm,angle=0}
\caption{Energy distribution of $\tilde\ell_R$ pair events per 
km$^2$, per year, at the detector. 
Curves that do not reach y axis; from top to bottom: 
$m_{\tilde q}=300$, ~$600$ and $900$~GeV. 
Here, $m_{\tilde\ell_R}=150$~GeV and
$m_{\tilde w}=250$~GeV. 
Also shown are the neutrino flux at earth and the $\mu$ and
the di-muon flux through the detector (curves that reach y axis; from top to bottom
respectively). In all cases we make use of the WB limit 
for the neutrino flux. 
}
\label{flux}
\end{flushright}
\end{figure}
In Figure~\ref{flux} we show the energy distribution for the NLSP 
pair events for three
choices of squark masses: $300$~GeV, $600$~GeV and $900$~GeV. Also 
shown are the neutrino flux at earth
in the WB limit, as well as the energy distribution of upgoing
$\mu$'s, and that of direct $\mu^+\mu^-$ production 
(see Section~\ref{back}). 
We see that, even for the heavier squarks, it is possible to 
obtain observable event rates. 
In Table~\ref{tab:nev} we show the event rates for $\tilde\ell_R$ 
pair production
per year and per km$^2$. The rates are given for the WB flux as well 
as for the 
Mannheim-Protheroe-Rachen (MPR) flux~\cite{mpr}, both for optically 
thin sources.  
For comparison, we also show the rates
for the $\mu^+\mu^-$ background. Thus, km$^3$
Cerenkov detectors such as 
IceCube, appear to be sensitive to most of the parameter space of
interest in scenarios with a relatively long lived NLSP.
\begin{table}
\begin{center}
\caption{Number of events per km$^2$ per year for different neutrino
fluxes at the Earth. The $\tilde{\ell}_R$ mass is 150 GeV and squark masses are
300, 600 and 900 GeV respectively. The signal events are integrated
from threshold ($E_\nu >1.6\times 10^5~$GeV), whereas the 
number of di-muon 
events are given for neutrino energies above $10^3$ GeV. 
The column $\mu^+\mu^-$ corresponds
to the di-muon background before track separation cuts are applied (see Section~\ref{back}).}
\begin{ruledtabular}
\begin{tabular}{l|cccc}
 & $\mu^+\mu^-$ & &$\tilde{\ell}_R\tilde{\bar\ell}_R$ & \\
~\\
& & (300) & (600) & (900)  \\
\hline\\
WB & 30 & 6 & 1 & 0.3 \\
MPR & 1412 & 21 & 3 & 1 
\label{tab:nev}
\end{tabular}
\end{ruledtabular}
\end{center}
\end{table}
The rates are comparable to the ones we obtained in Ref.\cite{abc}.

The NLSPs are produced in pairs very far from the detector 
and with a very large boost. Although the  angle between the two NLSP tracks,
is small due to the boost, the large range means that there will be 
a significant separation between the tracks. The question is then
whether this separation is large enough to be observed, but not too 
large so as to be larger than the dimensions of the detector.
A rough estimate can be made by defining the separation between NLSP
tracks as 
$\delta R\simeq L~\theta$, 
with $L$ the distance to the production point, typically  
a few $1000$~Km and 
$\theta\simeq p_{\rm SUSY}^{\rm CM}/p_{\rm boost}\simeq
(10^{-4}-10^{-5})$, the typical angle between the NLSPs in the lab frame. 
Then we should expect that the typical separation between the two 
NLSP tracks is of the order of
$\delta R\simeq {\rm few}\times (10-100){\rm m}$. 

We performed a more detailed calculation in order to  
have a more precise prediction for the 
distribution of the track separation of the events.
We consider the angular distribution of the two primaries in the center
of mass, as implied by eqns.~(\ref{llcc}) and (\ref{lrcc}) as well as 
those of the subdominant neutral exchange. 
We assume that the angular distribution of the right handed slepton 
pair in the 
center of mass is the same of that of the primary decay particles,  
the $\tilde{\ell}_L$ and $\tilde q$. 
This is a good approximation 
for energies well above threshold, where most events
come from, as it can be seen in Fig.~\ref{flux}. In this limit, it
is possible to define a ``thrust axis'', around which the decay
products of the two primary final states are boosted. 
We then boost this distribution to the laboratory frame.
The approximation made fails for energies just above threshold, 
for which decay products can have a more spherical distribution
resulting, in principle, in larger angles and ultimately in 
larger track separations. However, the events among these that do make
it to the detector tend to come from smaller distances than the range,
therefore compensating --at least in part-- the potentially larger angles.
In any case, this approximation tends to underestimate the stau track 
separation.
In the calculation we also include the additional zenith 
angular dependence induced by the earth profile. 

Taking the above features into consideration, we performed a simulation
where we generated approximately 30k events (the numbers for each squark
mass varied accordingly to the production energy threshold). These events
where distributed in energy steps ranging from the stau production threshold
to $10^{11}$ eV. For each energy we generated events for which 
the center of mass (CM) angular distribution was chosen based on the differential
cross section distribution, as mentioned above. 
The 4-momentum of these events were then defined in the 
CM and boosted to the lab frame. With this procedure we determined the lab angular
($\Theta_{\rm lab}$) distribution. Then the neutrino-Earth interaction
point (R) was chosen based on the interaction probability distribution. The separation
at the detector is then simply $\Theta_{\rm lab} \times$ R.

\begin{figure}
\centering
\epsfig{file=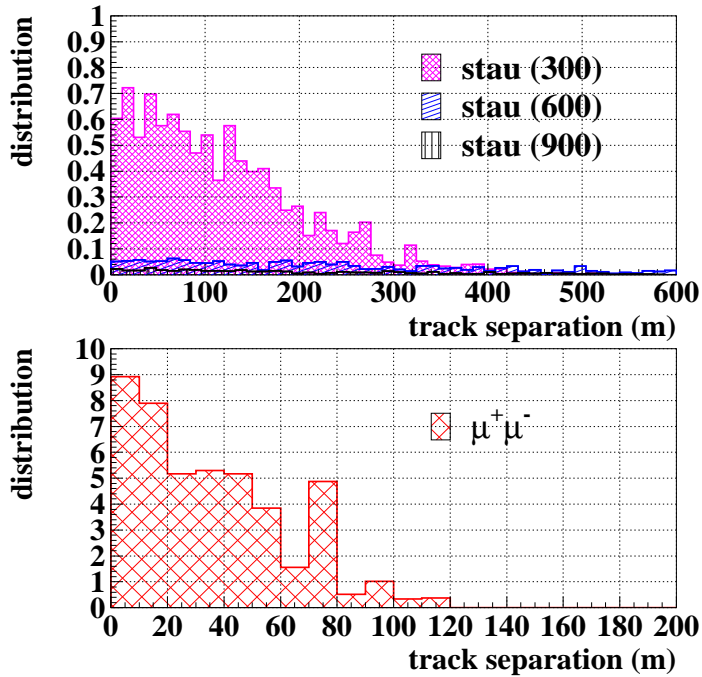,width=11.2cm,height=11cm,angle=0}
\caption{Top panel: Track separation distribution of $\tilde\ell_R$
pair events. 
Bottom panel: The track separation distribution of the di-muon
background.  The relative 
normalization corresponds to the relative number of events for signal 
and background. Note the different horizontal scales, as well as
different binning between the two figures.
}
\label{track_sep}
\end{figure}

The result is shown in Fig.~\ref{track_sep} shows the 
NLSPs track separation distribution for various squark masses. 
The rather broad distribution is characterized by track separations of 
hundreds of meters. 
For 300 GeV squark mass, 87\% of the distribution is above 30~m
separation, and 52\% is above 100~m separation. 

Another feature of the NLSP signal is that the two tracks
should have very small relative angles. The 
angular separation of the tracks in the signal is always well 
below a degree, the typical angular resolution 
of neutrino telescopes. so that the tracks will appear parallel to
each other.
Then, most NLSP events would consist of {\em two} parallel 
but well separated tracks, 
and are therefore expected to be very distinctive and different 
from backgrounds, as we show in Section~\ref{back}.  
Given that the track separation ranges up to (300-400)m, 
and that the typical dimension of the detector is $1~$Km, we expect that many
events will be contained and therefore distinguishable from
single-muon 
tracks. On the other hand, the typical separation appears 
to be large enough to distinguish the presence of two tracks in a
given signal event.

The energy with which the slepton pairs arrive at the detector 
has a distribution which peaks just above 100~TeV, as it can be seen 
in  Figure~\ref{endist}. 
On the other hand, the energy deposited in the detector by a typical
stau track ($\simeq 800$~m which is IceCube geometrical efficiency for long tracks
\cite{als}) is approximately $160~$GeV, which is small
compared to its total energy. Low energy muons may mimic this energy deposition pattern, 
but they tend to have shorter paths in the detector. This fact can be used in order
to reduce the background, as we discuss below.

\subsection{Backgrounds} 
\label{back}
The main potential background for the detection of NLSPs, 
muons produced by upgoing atmospheric neutrinos, is 
eliminated by asking for two tracks in the events. 
With the additional requirement that the tracks be almost parallel and
considering the excellent time resolution of the IceCube detector, 
the coincidence background (two coincident muons tracks) becomes 
negligible. 

However, there remains an important background, namely
direct di-muon production. The main source of direct di-muons
is the production of charm, which subsequently decays semileptonically
to a muon. That is, we consider the process
\be
\nu N\to \mu^- H_c \to \mu^- ~\mu^+ ~H_x~\nu  ~,
\nonumber
\ee
where the charm hadron $H_c$ decays according to $H_c\to H_x \mu^+\nu$,
and $H_x$ can be a strange or non-strange hadron.
The cross section for producing a charm quark from a light $d$ or $s$
quark is 
\bear
\frac{d^2\sigma(\nu N\to c X)}{d\xi dy} &=& \frac{G_F^2  2ME_\nu}{\pi}
\,R^2(Q^2)\,\xi\nonumber\\
&&\times\left(1-\frac{m_c^2}{2ME_\nu\xi}\right)\,\left\{
|V_{cs}|^2 s(\xi) \right.\nonumber\\
&&\left.+ |V_{cd}|^2 d(\xi)\right\}~,
\label{dimucs}
\eear
where $s(\xi)$ and $d(\xi)$ are the strange and down quark 
parton distribution functions respectively. In eqn.~(\ref{dimucs})
the slow scaling variable is defined as 
\be
\xi = x(1+\frac{m_c^2}{Q^2})~,
\label{slow}
\ee
and we also defined
\be
R(Q^2) = \frac{M_W^2}{M_W^2 + Q^2}~.
\label{rq2def}
\ee
The di-muon background results in a number of events
larger than the signal, even for the case of a 300~GeV squark mass, as it can be seen in 
Table~\ref{tab:nev}.
Also, the angular separation between the di-muon tracks is, as in the case
of the NLSP signal, very small and well below the projected 
angular resolution of neutrino telescopes. Thus, di-muon tracks will
also appear to be parallel.
\begin{center}
\begin{figure}[t]
%
\epsfig{file=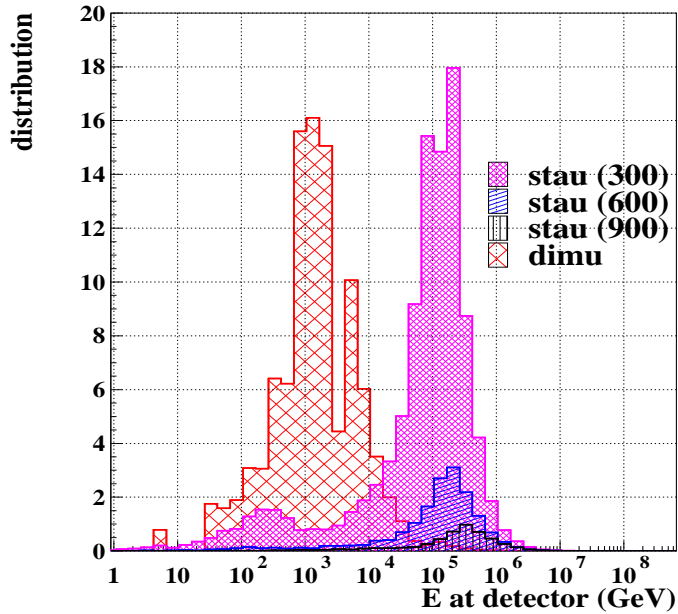,width=11.2cm,height=10cm,angle=0}
\caption{Arrival energy distribution 
of the $\tilde\ell_R$ at the 
detector and for $m_{\tilde q}=300,~600 {\rm ~and~} 900$~GeV..
Here, $m_{\tilde\ell_R}=150$~GeV and
$m_{\tilde w}=250$~GeV. 
Also shown is the arrival distribution for the di-muon
background. The energy {\em deposited} in the detector by a stau 
traveling the average track length of $800~$m is
$E_{\tilde{\ell}_R}^{\rm dep}= 160~$GeV, approximately the same for
all the masses considered here (See Figure~\ref{depos}).
}
\label{endist}
\end{figure}
\end{center}

\begin{figure}
%
\begin{center}
\epsfig{file=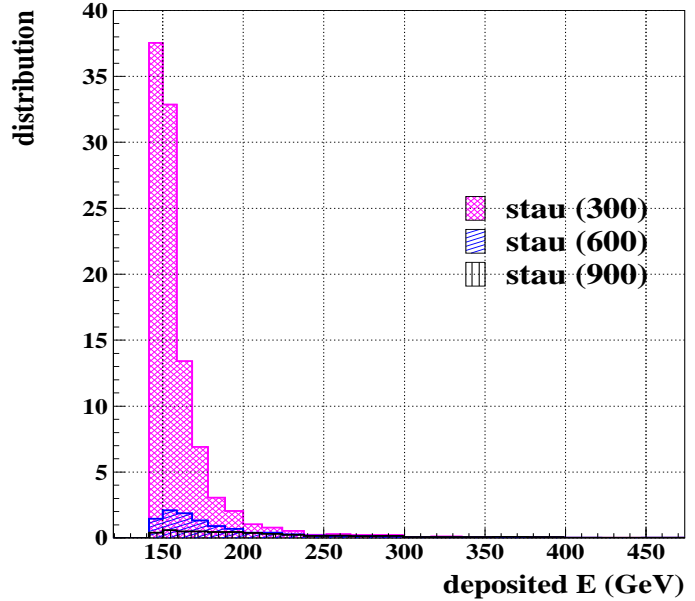,width=8.6cm,height=9cm,angle=0}
\caption{Energy deposited in the detector 
by a right-handed stau 
traveling the average track length of $800~$m. 
}
\label{depos}
\end{center}
\end{figure}

However, muons need to be close 
to the detector in order to range into it. Thus, 
the di-muon events have a considerably smaller separation between the two
tracks than the NLSP events, which can range in from hundreds or even 
thousands of kilometers. This is shown in Figure~\ref{track_sep},
where it is clear that for track separations above 100~m there should
not be any significant contribution from the di-muon background.
We can estimate the statistical significance of the signal as 
$S/\sqrt{B}$, with $S$ and $B$ the number of signal and background
events, respectively. We could then ask what is the cut in track
separation that would yield a significance equal to $5$. 
We find that cutting events with less than 106~m separation
results in 3 detectable NLSPs and 0.25 di-muons, yielding a  $5~\sigma$
significance.

It is possible to further reduce the di-muon background by making use of 
the energy deposition of the events.  
This is due to the fact 
that NLSPs go through the detector loosing very
little energy, while muons tend to deposit larger 
amounts of energy in the detector in such long paths. 
This can be seen by comparing the 
energy loss for a muon in Figure~\ref{bmu_fig} with that of a slepton
in Figure~\ref{bstau_fig}, where the largest source of energy loss for
the NLSP is still three orders of magnitude smaller than that for the 
muon at the energies with which they arrive at the detector.
For example, 
a muon with an arrival energy of $3\times 10^3~$GeV, transversing 
800 m of ice (the average through-detector length) 
will deposit $\sim 450$~ GeV in its path. This is to be compared with the deposited energy 
for a slepton in Figure~\ref{depos}. 
Thus, excluding events with {\em deposited} 
energies above a few hundred GeV, we can eliminate the high energy tail of the di-muon background 
of Figure~\ref{endist}. This can be used, in addition to the track separation cut,
to further reduce the background events. 
 The precise extent to which this method can be used to reduce the background 
requires further study. 


Finally, there is an additional potential source of di-muons from 
$\nu\gamma\to W^{+}\mu$ scattering~\cite{seckel}, 
where the $\gamma$ is emitted off the parton in the
nucleon and the produced $W^{\pm}$ decays to $\mu\nu_\mu$. 
However, not only this is a 
smaller source of di-muons, but also 
its track separation distribution is very similar to that of the
di-muons coming from 
charm production and
is equally eliminated by the same cut in track separation.

\section{Conclusions}
\label{conc}

We have confirmed that the proposal of Ref.~\cite{abc}, 
is viable, i.e. 
neutrino telescopes are
potentially sensitive to the
relatively long-lived charged NLSPs which are present in a wide
variety of models of supersymmetry breaking.  
Compared to Ref.~\cite{abc}, we have now made a detailed study of the 
NLSP energy loss (Section~\ref{eloss}). This has confirmed our 
estimates in Ref.~\cite{abc} and is in agreement with the 
calculations of Ref.~\cite{reno3}. 

Regarding the analysis of the signal, we have shown that 
the separation between tracks will be such that we can expect that most 
events will be contained in a ${\rm km}^3$ such as IceCube,
whereas they will be separated enough to be identified as two 
distinct tracks. 
Moreover, we have shown that  the separation between tracks is a 
variable that will allow to identify the signal above the 
most important background, the direct production of $\mu^+\mu^-$ 
pairs. This is shown in Figure~\ref{track_sep}. For instance, we have seem 
that a $5~\sigma$ significance can be achieved with a cut of 106~m in the 
track separation.

Also, given that the NLSP deposits energy in the detector like
a rather low energy muon, but it goes right through it like 
a high energy muon could, it is possible to further reduce the number of 
background events by making a cut in the amount of deposited energy.
of the events going through the detector. The details of this method of reducing the 
background deserves further study.

The region of the supersymmetry breaking parameter space that 
is available to neutrino telescopes is determined
by the requirement that the NLSP lifetime 
be long enough to give a
signal ($\sqrt{F}\agt 5 \times 10^6$~GeV).
Thus the observation of NLSP events at neutrino telescopes will 
constitute a direct probe of the scale of supersymmetry
breaking.  This is to be compared with the potential observation 
of these NLSPs at the Large
Hadron Collider (LHC), where for this range of $\sqrt{F}$,  
the NLSP decays outside the
detector and is seen through its ionization tracks. However, the 
observation at the LHC would
not constrain the NLSP lifetime significantly. 
Thus, we see that neutrino telescopes are
complementary to collider searches. For instance, the 
observation of NLSP events at the LHC, coupled to significantly fewer than
expected events in neutrino telescopes 
would point to $\sqrt{F}<10^{7}$~GeV.

The event rates shown in Table~\ref{tab:nev} are already encouraging
for experimental facilities that are been built, such as IceCube. 
Future upgrades of IceCube~\cite{iceplus} will result in even better 
sensitivity. The
water detectors ANTARES~\cite{antares}; NESTOR~\cite{nestor} and
NEMO~\cite{nemo} will be also be able to look for these NLSPs.

Finally, in the present work we focused on supersymmetry.  However,
other theories give rise to
relatively long lived charged particles which can be observed by neutrino
telescopes~\cite{abc2}. A large portion of this work could be applied
to these other scenarios with little modification.

{\it Acknowledgments ---} 
We thank M.~Alhers, J.~Kersten and A.~Ringwald for helpful
comments that lead us to find an error in an earlier version of this paper. 
I.~A. and G.~B. acknowledge the support of the State of S\~{a}o Paulo
Research Foundation (FAPESP). G.~B. also thanks the Brazilian  National Counsel
for Technological and Scientific Development (CNPq).
Z.~C. is supported by the NSF under grant PHY-0408954.


\begin{thebibliography}{99}


\bibitem{abc}
  I.~Albuquerque, G.~Burdman and Z.~Chacko,
  Phys.\ Rev.\ Lett.\  {\bf 92}, 221802 (2004).

\bibitem{GMSB}
M.~Dine, W.~Fischler, M.~Srednicki,
Nuc. \ Phys. \  B {\bf 189}, 575 (1981);
S.~Dimopoulos, S.~Raby,
Nuc. \ Phys. \  B {\bf 192}, 353 (1981);
L.~Alvarez--Gaum\'e, M.~Claudson, M.B.~Wise,
Nuc. \ Phys. \  B {\bf 207}, 96 (1982);
M. Dine, A.E. Nelson,
Phys. \ Rev. \ D {\bf 48}, 1277 (1993);
M. Dine, A.E. Nelson, Y. Shirman,
Phys. \ Rev. \ D {\bf 51}, 1362 (1995);
M. Dine, A.E. Nelson, Y. Nir, Y. Shirman,
Phys. \ Rev. \ D {\bf 53}, 2658 (1996).
For a review, see
G.F. Giudice, R. Rattazzi,
Phys.\ Rept.\  {\bf 322}, 419 (1999).

\bibitem{superWIMP}
J.~L.~Feng, A.~Rajaraman and F.~Takayama,
Phys.\ Rev.\ Lett.\  {\bf 91}, 011302 (2003)
[arXiv:hep-ph/0302215];
J.~L.~Feng, A.~Rajaraman and F.~Takayama,
Phys.\ Rev.\ D {\bf 68}, 063504 (2003)
[arXiv:hep-ph/0306024];
J.~L.~Feng, S.~f.~Su and F.~Takayama,
Phys.\ Rev.\ D {\bf 70}, 063514 (2004)
[arXiv:hep-ph/0404198];
J.~L.~Feng, S.~Su and F.~Takayama,
Phys.\ Rev.\ D {\bf 70}, 075019 (2004)
[arXiv:hep-ph/0404231].



\bibitem{icecube}
J.~Ahrens {\it et al.}  [The IceCube Collaboration],
Nucl.\ Phys.\ Proc.\ Suppl.\  {\bf 118}, 388 (2003).

\bibitem{bi}
  X.~J.~Bi, J.~X.~Wang, C.~Zhang and X.~m.~Zhang,
  Phys.\ Rev.\ D {\bf 70}, 123512 (2004)
  [arXiv:hep-ph/0404263].
                                                                                           
\bibitem{reno3}
  M.~H.~Reno, I.~Sarcevic and S.~Su,
  Astropart.\ Phys.\  {\bf 24}, 107 (2005)
  [arXiv:hep-ph/0503030].
                                                                                           
\bibitem{ring}M.~Ahlers, J.~Kersten and A.~Ringwald,
  arXiv:hep-ph/0604188.


\bibitem{tau1}
  S.~I.~Dutta, M.~H.~Reno, I.~Sarcevic and D.~Seckel,
  Phys.\ Rev.\ D {\bf 63}, 094020 (2001).

\bibitem{tau2}
  E.~V.~Bugaev and Y.~V.~Shlepin,
  Phys.\ Rev.\ D {\bf 67}, 034027 (2003).


\bibitem{als} For a detailed discussion of neutrino event rates see
I.~Albuquerque, J.~Lamoureux and G.~F.~Smoot,
Astrophys.\ J.\ Suppl.\  {\bf 141}, 195 (2002).

\bibitem{pdg}
 S.~Eidelman {\it et al.}  [Particle Data Group],
  Phys.\ Lett.\ B {\bf 592} (2004) 1.

\bibitem{groom}
D.~E.~Groom, N.~V.~Mokhov and S.~I.~Striganov, Atomic Data and 
Nuclear Data Tables {\bf 78}, 183 (2001).


\bibitem{kelner}
  S.~R.~Kelner and Y.~D.~Kotov,
 Sov.\ J.\ Nucl.\ Phys.\ 
  (Yad.\ Fiz.\  {\bf 7} 360 (1968)).

\bibitem{kokoulin} 
R.~P.~Kokoulin and A.~A.~Petrukhin, Acta Physica
Acad.\ Sci.\ Hung., {\bf 29}, Suppl.~{\bf 4}, 277 (1970).


\bibitem{allm1}
  H.~Abramowicz, E.~M.~Levin, A.~Levy and U.~Maor,
  Phys.\ Lett.\ B {\bf 269}, 465 (1991).

\bibitem{allm2}
  H.~Abramowicz and A.~Levy,
  arXiv:hep-ph/9712415.

 
\bibitem{wb}
E.~Waxman and J.~N.~Bahcall,
Phys.\ Rev.\ D {\bf 59}, 023002 (1999); 
J.~N.~Bahcall and E.~Waxman,
Phys.\ Rev.\ D {\bf 64}, 023002 (2001).

\bibitem{mpr}
K.~Mannheim, R.~J.~Protheroe and J.~P.~Rachen,
Phys.\ Rev.\ D {\bf 63}, 023003 (2001)

\bibitem{gqrs}
R.~Gandhi, C.~Quigg, M.~H.~Reno and I.~Sarcevic,
Astropart.\ Phys.\  {\bf 5}, 81 (1996); 
R.~Gandhi, C.~Quigg, M.~H.~Reno and I.~Sarcevic,
Phys.\ Rev.\ D {\bf 58}, 093009 (1998). 

\bibitem{seckel} D.~Seckel,
  Phys.\ Rev.\ Lett.\  {\bf 80}, 900 (1998)
  [arXiv:hep-ph/9709290].

\bibitem{iceplus}
F.~Halzen and D.~Hooper,
arXiv:astro-ph/0310152.

\bibitem{antares}J.~A.~Aguilar {\it et al.}  [The ANTARES Collaboration],
arXiv:astro-ph/0310130.

\bibitem{nestor}P.~K.~F.~Grieder  [NESTOR Collaboration],
Nuovo Cim.\  {\bf 24C} (2001) 771.

\bibitem{nemo} NEMO Collaboration, http://nemoweb.lns.infn.it/

\bibitem{abc2}
I.~Albuquerque, G.~Burdman, C.~Krenke, and B.~Nosratpour, in preparation.

\end{thebibliography}
\end{document}